# Characterization of turbulent supersonic flow over a backward facing step through POD

R. K. Soni[*], N. Arya[†], Ashoke De[‡]

Indian Institute of Technology Kanpur, Kanpur -208016, Uttar Pradesh, India

**ABSTRACT**

The present work reports on the flow physics of turbulent supersonic flow over backward facing step (BFS) at Mach 2 using LES methodology where the dynamic Smagorinsky model is used for SGS modeling, while POD is invoked to identify the coherent structures present in the flow. The mean data obtained through the computations is in good agreement with the experimental measurements, while the iso-surfaces of Q-criterion at different time instants show the complex flow structures. The presence of counter rotating vortex pair in the shear layer along with the complex shock wave/boundary layer interaction leading to the separation of boundary layer is also evident from the contours of both Q and the modulus of vorticity. Further, the POD analysis reveals the presence of coherent structures, where the first and second modes confirm the vortical structures near the step as well as along the shear layer in the downstream region; while the second, third and fourth modes confirm the presence of vortices along the shear layer due to Kelvin-Helmholtz (K-H) instability. Moreover, POD as well as frequency analysis is extended at different planes to extract the detailed flow features.

[*] Graduate student, Department of Aerospace Engineering
[†] Graduate student, Department of Aerospace Engineering
[‡] Associate Professor, Department of Aerospace Engineering
[‡]Corresponding Author: Tel.: +91-51-2597863 Fax: +91-512-2597561; E-mail address: ashoke@iitk.ac.in

1. **INTRODUCTION**

The supersonic flow past BFS is a classical problem, is widely studied for different applications ranging from supersonic mixing, flame stabilization, ballistics to space shuttle and many more. A thorough understanding of the complex flow physics encountered in supersonic BFS can directly benefit many of these design and developments. Some of the flow features includes flow separation, reattachment and viscous-inviscid interactions. In various experimental observations [1-7], it is found that the flow separates slightly below the corner at the step through the formation of a lip shock. The free stream above the corner undergoes an expansion, while the pressure decreases after the step. The separated boundary layer behind the step develops into a free shear layer in a region of constant pressure. On approaching the wall, the shear layer is compressed at the downstream of the step and re-attaches through the formation of an oblique shock. In the base region, the pressure remains relatively uniform but slightly lesser compared to the free stream.

After the reattachment point, the streamlines become parallel to the walls and the flow tends to behave like a supersonic flow over a flat plate. The dynamic differences between subsonic and supersonic boundary layers can be explained by considering the variations in the properties of fluid due to the temperature variation. Morkovin [8] proposed that as long as the fluctuating Mach number remains small, the compressible boundary layers follow the same dynamics as the incompressible boundary layers. This hypothesis is the basis for the van Driest transformation which allows the comparison of compressible boundary layer profile with the incompressible distribution. Many researchers [9-12] have used the van Driest transformation [13] to study the turbulent supersonic boundary layer profiles and have established the validity of the Morkovin's hypothesis. The validation of the wall model for the present work is also based on van Driest transformation with Prandtl number (Pr) = 0.75 as used by Fernholz & Finley [9] contrary to van Driest who used Pr = 1.

Troutt et al. [14] in their study found the large scale structures near the reattachment region. The increment of the flow residence time due to the presence of the step was reported by various authors [15]. Huang et al. [16] observed that the vortices generated at the step corner and in the base region enhance the fuel and air mixing which is reflected in the improved combustion and mixing efficiency. This can be attributed to the fact that the generation of recirculation region with relatively low flow velocity allows fuel and air to co-exist for prolonged time, which is highly desirable phenomenon for supersonic combustion.

It is quite well known that the coherent structures are fundamental to the understanding of turbulent flows [17, 18]. Moreover, this understanding is also very important for many engineering problems involving noise, mixing, drag and others, because they contain large portion of the fluctuating energy. Therefore, it is well known that the turbulent flows can be better analyzed and understood as well, by studying the dynamics of coherent structures. The proper Orthogonal Decomposition (POD), also known as Karhunen-Loeve expansion, and the principle component analysis has been utilized in various areas such as image processing [19], turbulent flows [20], pattern recognition [21] and structural vibrations [22, 23]. Lumley [24] introduced the POD technique to study turbulent flows by capturing the most energetic flow structures. Over the years, it has emerged as a powerful tool for the extraction of dominant structures in turbulent flows. POD has been widely used to construct low-dimensional model to study the dynamics of unsteady flow field. There exists various class of POD techniques of which, direct method and method of snapshots are widely used. In the present study, the method of snapshots proposed by Sirovich [25] has been invoked to study the dominant modes of the flow field. For the large scale non-linear problems, method of snapshots has been found to be more economical than the direct method [26].

Numerical simulation of flows with shock and other complex phenomena like shock-shock interaction etc. are inherently challenging both in terms of the modeling and the numerical implementation. These unsteady flows cannot be captured accurately by Reynolds-averaged Navier-Stokes (RANS) formulations, and require the use of Large Eddy Simulation (LES) methodology. LES offers resolution of all large scales however modeling is required to capture the small scales (sub-grid scales). A cut-off length scale termed as the filter width is defined to distinguish the resolved scales from the modeled scales.

In the past, various numerical studies [1, 15-16] focused on characterizing the BFS flow field through RANS, but with limited success. In most of the cases, mean data prediction was found to be acceptable but the complete understanding of various phenomena like K-H type instability, shear layer oscillation and dominating structures could not be characterized. Hence, the present work is directed towards the understanding of flow physics in a supersonic backward facing step, which to the best of authors' knowledge is sparsely reported in open literature. Majority of the numerical investigations is focused on understanding the mean flow features through RANS, which cannot provide the detailed understanding of the unsteady flow field. Therefore, the main objective of the present work is to perform LES along with POD analysis to shed light on the understanding of detailed flow features in a supersonic flow past backward facing step. It is expected that the unsteady analysis of the LES results would offer better understanding of the evolution of flow structures along with the presence of various dominating structures. The POD analysis, which is also not reported in the open literature for this kind of flow, is also used to identify different flow structures present in the flow field considered herein. Furthermore, the POD analysis is extended to various planes in order to characterize the dominating and transient flow structures present in the domain. In addition, the POD temporal coefficients at different planes are analyzed to account for the pressure fluctuations in the domain. In the first part of the study, LES results including validation are discussed followed by the detailed POD analysis in the later part.

## 2. NUMERICAL DETAILS

### 2.1. Governing Equations

The filtered governing equations for the conservation of mass, momentum, and energy are solved and recast as:

Continuity equation:

$$\frac{\partial}{\partial t}(\bar{\rho}) + \frac{\partial}{\partial x_i}(\bar{\rho}\tilde{u}_i) = 0 \quad (1)$$

Momentum equation:

$$\frac{\partial}{\partial t}(\bar{\rho}\tilde{u}_i) + \frac{\partial}{\partial x_j}(\bar{\rho}\tilde{u}_i\tilde{u}_j) = -\frac{\partial}{\partial x_i}(\bar{p}) + \frac{\partial}{\partial x_j}\left((\mu+\mu_t)\frac{\partial \tilde{u}_i}{\partial x_j}\right) \quad (2)$$

Energy equation:

$$\frac{\partial}{\partial t}(\bar{\rho}\tilde{E}) + \frac{\partial}{\partial x_i}(\bar{\rho}\tilde{u}_i\tilde{E}) = -\frac{\partial}{\partial x_j}\left(\tilde{u}_j\left(-\tilde{p}I + \mu\frac{\partial \tilde{u}_i}{\partial x_j}\right)\right) + \frac{\partial}{\partial x_i}\left(\left(k + \frac{\mu_t C_p}{\Pr_t}\right)\frac{\partial \tilde{T}}{\partial x_i}\right) \quad (3)$$

Where (~) and (-) in above equation refer to filtered and Favre averaged quantity, $\rho$ is the density, $u_i$ is the velocity vector, p is the pressure, $E = e + u^2_i /2$ is the total energy, where $e = h - p/\rho$ is the internal energy and h is enthalpy. The fluid properties $\mu$ and $k$ are respectively the viscosity, and the thermal conductivity, while $\mu_t$ and $\Pr_t$ are the turbulent eddy viscosity, and the turbulent Prandtl number respectively.

### 2.2. Dynamic LES model and wall function modelling

To model the turbulent eddy viscosity, LES is used so that the energetic larger-scale motions are resolved, and only the small scale fluctuations are modeled. Standard Smagorinsky model performs poorly near massively separating fields and near walls [27-29] which makes the dynamic Smagorinsky model more feasible for the present geometry which involves flow separation at the step wall followed by the reattachment at the lower wall. Yoshizawa [30] proposed an eddy viscosity model which uses the Smagorinsky model to account for the anisotropic part of the SGS stresses while the SGS energy was modeled separately. However, Moin et al. [31] proposed a modification to this model by determining the two model coefficients dynamically, rather than an apriori input. They utilized Germano identity [32], which relates the SGS stresses $\tau_{ij}$ to the resolved turbulent stresses (Leonard stresses) $L_{ij}$ and sub test stresses $T_{ij}$. A dynamic model based on the formulation utilized by Martin et al. [33] which was proposed by Lilly [34] is invoked in the present study. The Smagorinsky coefficients are calculated as:

$$\tau_{ij} - \frac{\delta_{ij}}{3} = -C_s^2 2\bar{\Delta}^2 \bar{\rho} \left( \tilde{S}_{ij} - \frac{\delta_{ij}}{3} \tilde{S}_{kk} \right) = C_s^2 \alpha_{ij} \qquad (4)$$

$$\tau_{kk} = C_I 2\bar{\rho}\bar{\Delta}^2 \left| \tilde{S}_{ij} \right| \qquad (5)$$

where, $\left| \tilde{S} \right| = \left( 2\tilde{S}_{ij} \tilde{S}_{ij} \right)^{1/2}$

$$C = C_s^2 = \frac{\langle L_{ij} M_{ij} \rangle}{\langle M_{kl} M_{kl} \rangle} \qquad \text{and} \qquad C_I = \frac{\langle L_{kk} \rangle}{\langle \beta - \alpha \rangle} \qquad (6)$$

where,

$$\left. \begin{array}{l} \beta_{ij} = -2\hat{\Delta}^2 \hat{\bar{\rho}} \left| \breve{\tilde{S}} \right| \left( \breve{\tilde{S}}_{ij} - \delta_{ij} \breve{\tilde{S}}_{kk}/3 \right) \\ M_{ij} = \beta_{ij} - \hat{\alpha}_{ij} \\ \beta = 2\hat{\bar{\Delta}}^2 \hat{\bar{\rho}} \left| \breve{\tilde{S}} \right|^2 \end{array} \right\} \qquad (7)$$

The model coefficients that have been implemented in the present work are calculated using averaging at the cell faces (local averaging) as opposed to the global averaging for the calculation. The global averaging for supersonic case which is associated with shock waves and expansion fans is highly unfeasible since the calculated Smagorinsky constant creates numerical instability. The dynamic model is clipped in such a way so that the back-scattering is allowed. It is clipped as $\left( -\mu_{molecular} \right) < \mu_{sgs}$.

Since the main aim of the study is to characterize the complete flow field past a backward facing step, boundary layer has not been resolved but modeled using a wall model to reduce the computational cost of the simulations. The modelling of the boundary layer is achieved through a wall model which uses Spalding's law of the wall to create a smooth profile for $\mu_{sgs}$ from the wall until the first grid point. The Spalding's law of the wall is given by

$$y^+ = u^+ + 0.1108(e^{0.4u^+} - 1 - 0.4u^+) \tag{8}$$

The frictional velocity and shear stress at the wall are given by the relation as:

$$u_\tau = \sqrt{\tau_w/\rho} \tag{9}$$

$$\tau_w = (\mu_{mol} + \mu_{sgs})\frac{\partial u}{\partial y} \tag{10}$$

$u^+$ and frictional velocity are related as:

$$u^+ = U*\sqrt{\rho/\tau} \tag{11}$$

The $\mu_{sgs}$ and $\mu_{eff}$ from the Spalding's law of the wall follow the relation

$$\frac{\mu_{sgs}}{\mu_{eff}} = 1/[1 + \frac{1}{0.04432}\{e^{0.4u^+} - 1 - 0.4u^+ - \frac{(0.4u^+)^2}{2!}\}] \tag{12}$$

$u_\tau$ is solved iteratively from the eqn. 9-11 and using eqn. 12, the value of $\mu_{sgs}$ is calculated. Thus a smooth profile for $\mu_{sgs}$ is created by the wall function. The wall function also clips the $\mu_{sgs}$ values in order to avoid numerical instability. The wall model employed in the present LES simulations is validated against the DNS results of supersonic flow over a flat plate [11, 12] in order to gain confidence on this wall-modelling approach. The non-dimensional wall distance is plotted against transformed van Driest velocity which is presented in the Figure 1. Though, there is a slight deviation in the prediction of the wall model from the DNS values in the outer boundary layer, the value predicted by the wall model for y+ less than 100 is in excellent agreement with the DNS results. Thus, it is quite evident that the application of the wall model in LES provides good results with a considerable reduction in the computational cost.

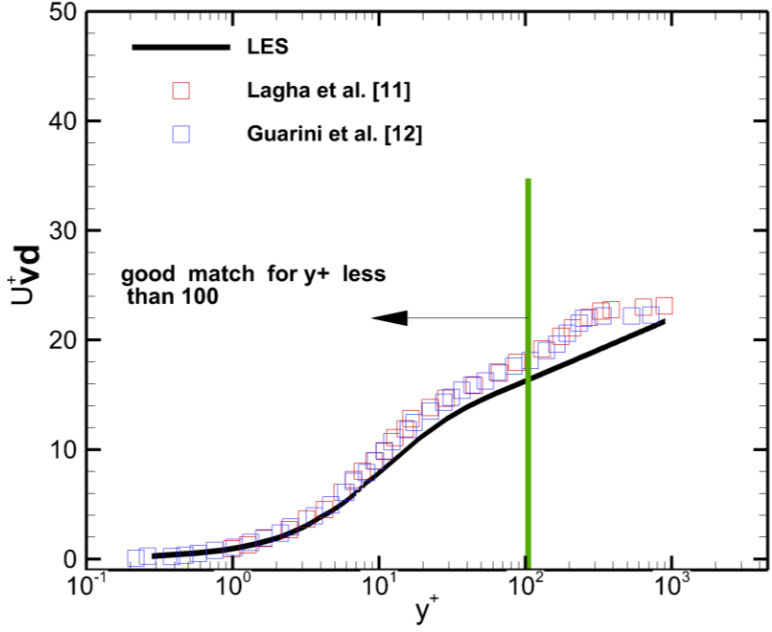

**Figure 1**: Van Driest transformed velocity

**2.3. Method of Snapshots**

POD was introduced in fluid mechanics by Lumley [24] to characterize the behavior of turbulent flows. The present analysis is done using the computationally less expensive "method of snapshots" by Sirovich [19, 25]. Let $U_i^N = U(x, t_N)_i$ where N represents the number of snapshots taken from a simulation or an experiment for a variable and the index *i* which runs from 1 to *M* represents the spatial location where the values are recorded. The basis functions capture the most energetic structures (coherent structures) in a turbulent flow, the choice of the variables for POD analysis is important. For an incompressible flow, the total energy is more or less represented by the total kinetic energy of the system and consequently only the velocity vectors are of importance. On the other hand, in a compressible flow the internal energy contributes a substantial part of the total energy. Therefore, a thermodynamic variable should be included in the analysis. For the present case, three velocity components, density and temperature is used for the analysis.

$$U_i^N = U(u, v, w, T)_i^N \qquad (13)$$

These components are obtained by subtracting the mean flow field from the instantaneous one. Since the vector *U* contains velocity as well as thermodynamic variables, the standard inner product is no longer valid for the analysis. A separate inner product is defined as,

$$(x; y) = \int_\Omega \left( x_1 y_1 + x_2 y_2 + x_3 y_3 + \gamma x_4 y_4 \right) ds \tag{14}$$

Here scaling factor γ is used which is necessary to balance the velocity and temperature fluctuation energies, so that the most energetic fluctuations of both the temperature and the velocity fields are captured efficiently by the basis functions.

Lumley and Poje [27] also calculated the optimum value of γ that maximizes the average of the square of the projection of data on to the basis functions. The value of γ is defined as [26],

$$\gamma = \frac{\int_\Omega (\rho\rho + uu + vv + ww) ds}{\int_\Omega (TT) ds} \tag{15}$$

Thus, the obtained basis functions $\phi = (\phi_1, \phi_2, \phi_3, \phi_4)$ follow the integral eigenvalue problem,

1.
$$\int_\Omega C_{ij}(x, x') \phi_j(x) ds = \lambda \phi_i(x) \tag{16}$$

where $i$ and $j$ varies from 1 to 5 and $C_{ij}$ is an auto co-variance matrix created as,

2.
$$C_{ij}(x, x') = \left( U_i(x,t); U_j(x',t) \right) \tag{17}$$

Now the eigenvectors obtained are normalized by using a standard norm definition given by,

3.
$$\|X\| = \sqrt{\left( X_1^2 + X_2^2 + X_3^2 + \dots X_N^2 \right)} \tag{18}$$

The solutions are ordered in decreasing order of the eigenvalues. The highest eigenvalue corresponds to the highest energy mode.

$$\lambda^1 > \lambda^2 > \lambda^3 \dots \lambda^N \tag{19}$$

The eigenvectors corresponding to the eigenvalues make up the basis for construction of POD modes. Once we have the POD modes $\phi_i$ the flow field can be represented as

4.
$$U(x,t) = \Sigma_1^N \phi_i(x) a^i(t) \tag{20}$$

Fukanaga [21] in their study have shown that the total kinetic energy content from the velocity fluctuations in the snapshots for a given POD mode is in proportion to the corresponding eigenvalue. It is because of this,

5.

eigenvalues in eq. (13) are ordered in such fashion to make sure that the most energy containing modes are the first ones. Thus, the dominant flow structures are captured in the first few POD modes.

**Table 1:** Domain dimension and grid distribution details

|  | $L_1$ (mm) | $L_2$ (mm) | $L_3$ (mm) | $L_4$ (mm) | $N_x$ | $N_y$ | $N_z$ |
|---|---|---|---|---|---|---|---|
| Coarse (1.62 M) | 3.2 | 80 | 25 | 21.29 | 270 | 120 | 40 |
| Fine (5.01 M) | 3.2 | 80 | 25 | 21.29 | 380 | 210 | 80 |

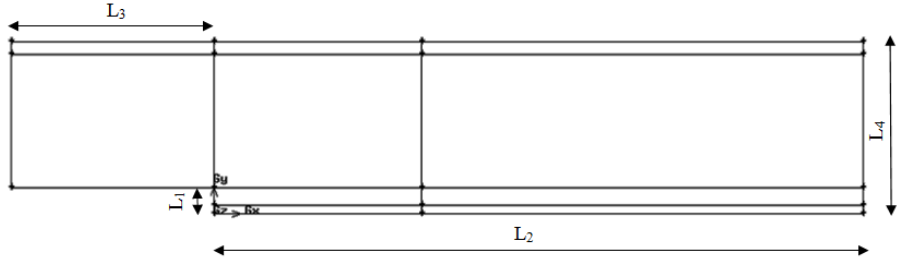

**Figure 2**: Details of computational domain with multi-block structure

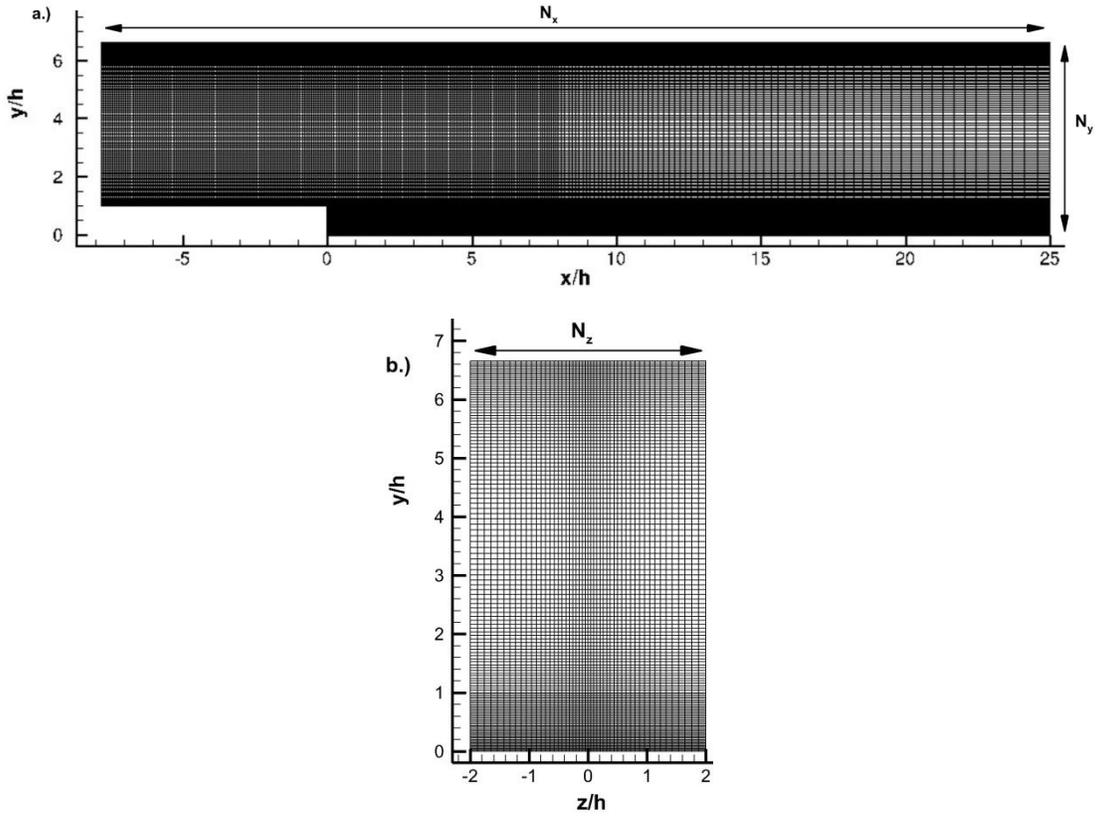

**Figure 3:** Description of coarse grid distribution in a) xy and b) zy plane

## 2.4. Computational details

Figure 2 presents the computational domain with multi-block structure to accommodate varying grid size. The domain is discretized using non-uniform hexahedral cells. Figure 3 exhibits the grid distribution in *xy* and *zy* plane. It can be noticed that more grid points are clustered in the region prone to high pressure gradient and small scales. Two sets of grids have been generated with 1.62 and 5.01 million cells, respectively. For both the grids, the bi-exponential stretching is used in spanwise direction, while uniform grid is distributed along the approach length as well as post step region till $x/h = 8$; after which the exponential stretching along the streamwise direction is performed. Along the wall normal direction, stretching is utilized from the lower wall up to the corner of the step with bi-exponential stretching between the step and the top wall. The main objective while generating grid is the ability to resolve major flow features experienced for a supersonic flow past backward facing step.

The filter size is taken care of during the grid generation. Using the theoretical calculation of the inertial and the dissipation length scales, the filter size is decided to make sure that the filter width lies within the inertial and dissipation sub range. A post simulation analysis is carried out using the Discrete Fourier Transform of the time evolution of pressure for two different grids as presented in the Figure 4. Furthermore, the quality of the grid is analyzed with the help of $LES_{IQ}$ criteria as proposed by Pope [35], which is the ratio of the resolved kinetic energy to the total kinetic energy $LES_{IQ} = k_{resolved}/k_{total}$ ; where $k_{resolved}$ is proportional to the $u_{rms}^2$ and $k_{total}$ is the sum of the resolved kinetic energy and the SGS kinetic energy. Typically, for a good LES the ratio of these two quantities should be more than 0.8. It is evident from the Figure 5 that the chosen grid is able to resolve the maximum energy in the domain within fairly acceptable accuracy. To utilize the wall function, $y^+$ value of 40 and 20 is used for the coarse and the fine grid, respectively. Information regarding the domain dimension and grid is presented in Table 1.

At the inlet boundary, uniform flow properties such as static pressure ($P_\infty$ = 35KPa) and static temperature ($T_\infty$ = 167 K) are specified providing Mach 2 flow, Re=1.024×10$^5$ whereas turbulent inflow with 5% fluctation is applied to velocity field. No-slip boundary conditions are enforced at the top and bottom wall along with condition that normal pressure gradient vanishes at wall. At the outlet, flow variables are extrapolated and non-reflecting boundary condition (NRBC) is imposed to allow outgoing waves to exit flow domain without reflection [36]. This boundary condition provides a wave transmissive outflow condition based on solving $\frac{\partial}{\partial t}(\psi, U) = 0$ at the boundary.

The wave speed is calculated using $x_p = \frac{\varphi_p}{|s_f|} + \sqrt{\frac{\gamma}{\psi_p}}$; where $x_p$ is the value of the field at the patch, $\varphi_p$ is the flux value of the field; $s_f$ is the face area vector and $\psi_p$ is the patch compressibility. Numerical results are obtained by employing the dynamic SGS model inside the density based solver (rhoCentralFoam) in OpenFOAM framework [37, 38]. The dynamics SGS model utilized for the present computation is based on dynamic calculation of two model constants. Second order backward Euler scheme is used for temporal discretization while the convection and diffusion terms are discretized using second order low dissipation filtered-linear scheme [37] and central difference scheme, respectively. The simulation is carried out for 70 non-dimensional times with time stepping of $10^{-8}$ that allows the CFL number to be maintained below 0.5.

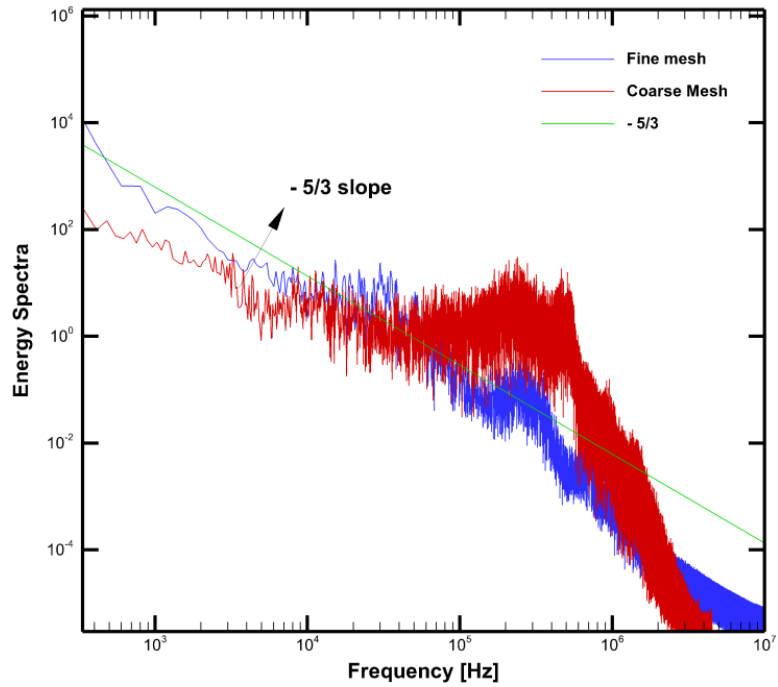

**Figure 4**: Power Spectral Density of pressure fluctuation

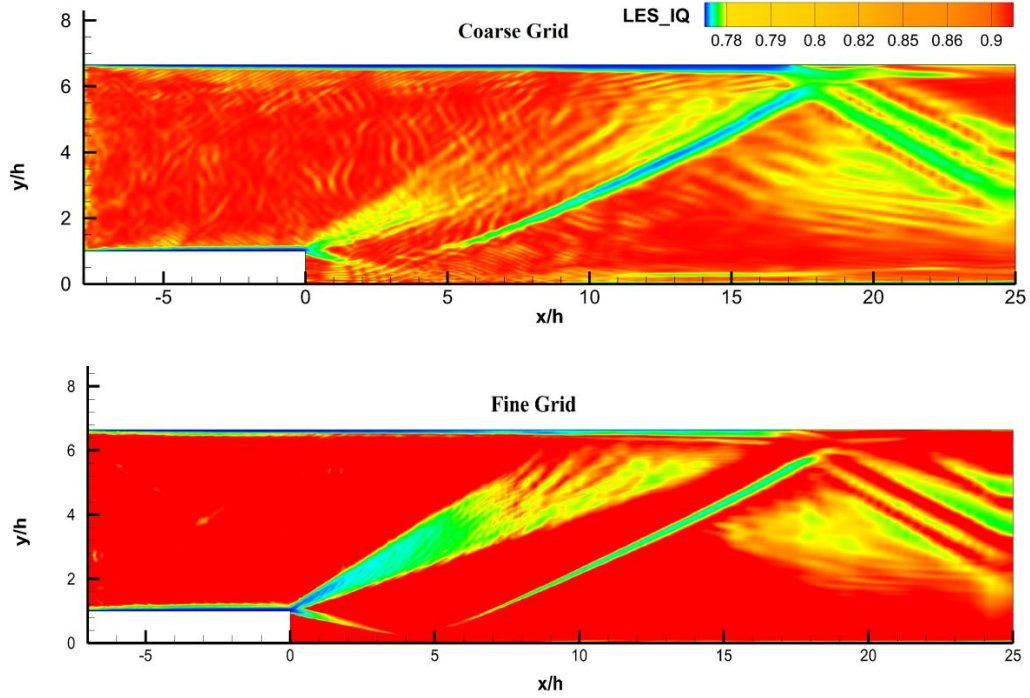

**Figure 5:** Ratio of resolved to total turbulent kinetic energy

## 3. RESULTS AND DISCUSSION

### 3.1: Mean flow field and visualization

The numerical results computed through LES are validated against the experimental results. The streamwise velocity and pressure profiles normalized with free stream values at three locations, namely, $x/h$ = 1.75, $x/h$ = 3 and $x/h$ = 6.66, are compared with the experimental data of McDaniel et al. [7] and presented in the Figure 6. It is observed that the numerically predicted results match well with the experimental data except in the region of $y/h$ < 1.

The discrepancy in the pressure profile at $x/h$ = 6.66 in near wall region is higher as compared to $x/h$ = 1.75. This is due to the fact that the former location is after the re-attachment shock which is in non-isentropic region. However, streamwise velocity matches quite well with the experimental observation.

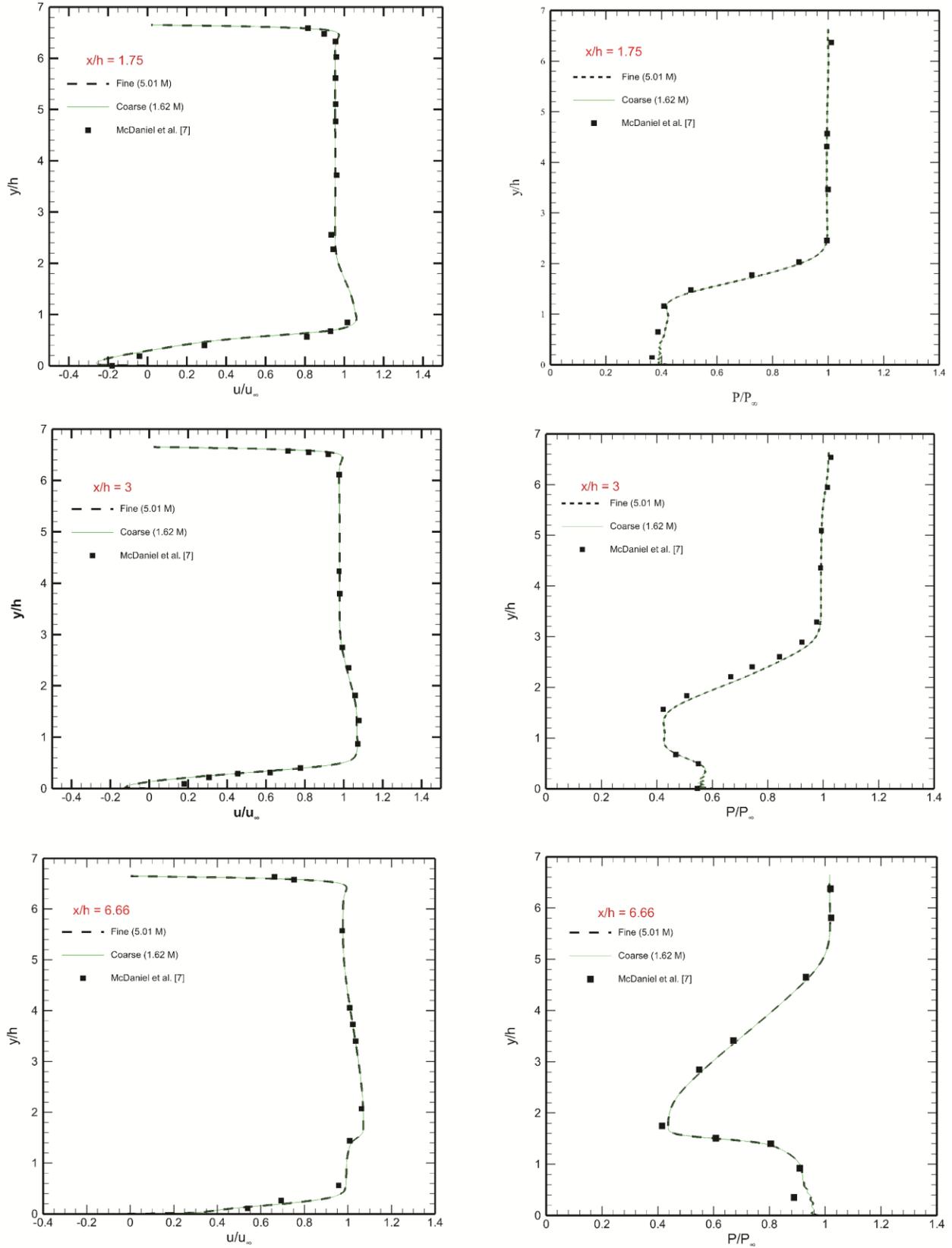

**Figure 6:** Non-dimensional pressure and streamwise velocity plot at different axial locations

It can be noticed that in the experimental observation at *x/h* = 6.66, pressure is slightly lower than the free stream pressure, which slowly increases up to slightly above the step height. However, numerical result appears to be slightly over predicted in the near wall region and up to the step height with values approximately close to the free stream pressure. Monotonic increase in pressure across expansion fan can be seen from the experimental results, and in this region the numerical predictions appear to be in excellent agreement; while the pressure remains constant above the expansion fan. In addition, the results using two different grids are found to be in excellent agreement with each other. Hence, the further detailed discussion is carried out using the coarse grid results only, unless or otherwise specified.

From the Figure 6, it is established that the quantitative agreement with experimental results along the centerline plane is in good agreement; however contour plots are also important to identify the major flow features. Figure 7 (b) presents the contour plots of normalized streamwise velocity and pressure along the centerline plane. It can be seen that the observation is consistent with the various experimental studies performed in the past [1-4]. It is clearly visible that the shear layer, re-circulation zone and the reattachment shock are accurately captured.

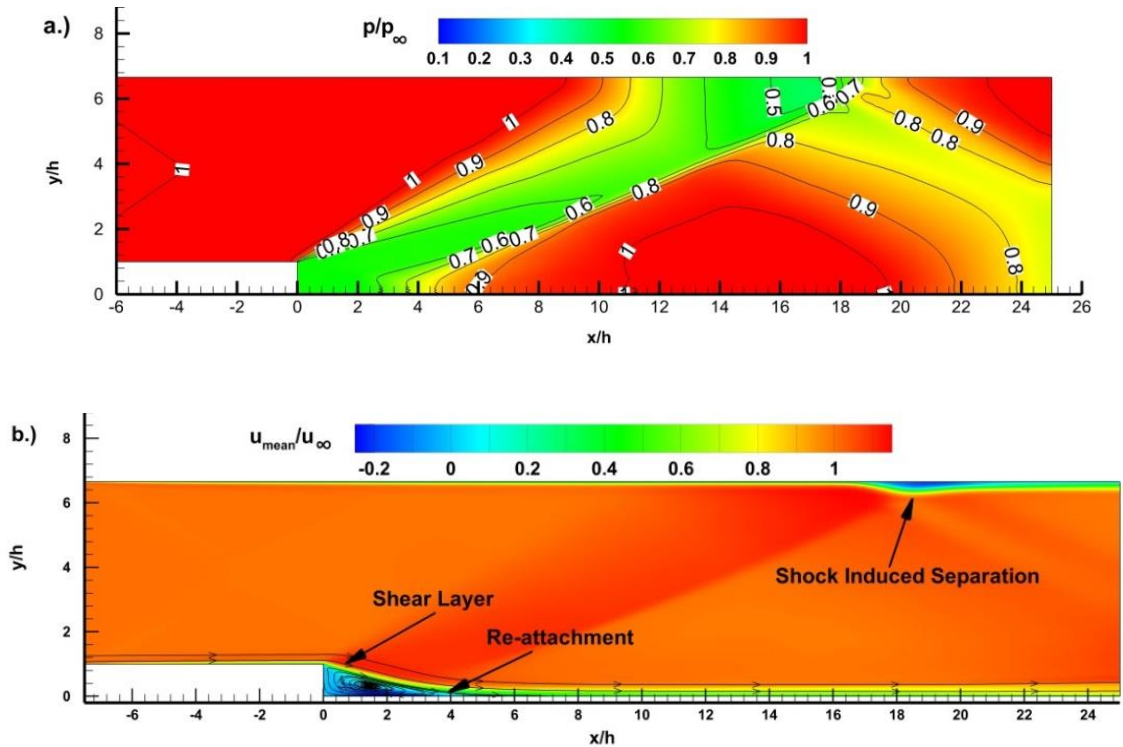

**Figure 7:** a) Normalized pressure contour, b) Normalized streamwise velocity contour along centerline plane

The pressure contour lines along the expansion fan are essentially straight line as can be observed from Figure 7(a). It can be witnessed that the pressure in the base region is almost uniform but slightly lesser than the pressure immediately downstream of the expansion fan. It is apparent that initially the contour lines are separate which coalesce further downstream to form well-defined re-attachment shock. It is important to note that the pressure remains quiet uniform across the boundary layer, however small pressure gradient is observed across the shear layer. It can be verified from Figure 7(b) that the flow does not separate immediately at the step but separates slightly below the step which is consistent with the observation of [4-5], through the formation of lip shock.

One of the most important parameters in backward facing step is the reattachment length which can be estimated through various parameters. The reattachment length as predicted by present computations is $x/h \approx 4.1$ for both the grids, which (in case of coarse grid) can be verified from the Figure 7(b). Although the recirculation length is provided, a comparison is not possible as literature does not provide the estimate from the experimental observation.

In Figure 8, the modulus of vorticity is presented for various instants to present the shear layer oscillations and evolution of the flow field. As observed initially, the large transient structure near the step is formed which later develops into the shear layer. Further, through $t \approx 0.0012s$ - $0.0034s$, the shear layer slowly develops and upon reaching the wall at $x/h \approx 2$-$3$ is compressed and re-attaches through the formation of oblique re-attachment shock wave. Beyond this point, the shear layer is developed into boundary layer. Also, it can be observed that due to the impingement of re-attachment shock wave on the upper wall ($x/h \approx 18$-$20$) the vorticity increases due to baroclinic torque effect. The impingement of shock wave leads to the vortex shedding phenomenon downstream of this point. The contour plot reveals that at this point of impingement, flow is slightly shifted away from the wall which suggests that the streamlines are curved at this point, which is due to the formation of local re-circulation region. It is clearly visible that the shear layer starts to roll around, between $t \approx 0.0012s$ to $t \approx 0.0034s$, leading to the formation of structures similar to counter rotating vortex pair (CVP), as also reported by Brown and Roshko [17]. However, upon reattaching, this CVP ceases to exist with distinctly visible K-H instability, which is apparent from the Figure 8(d).

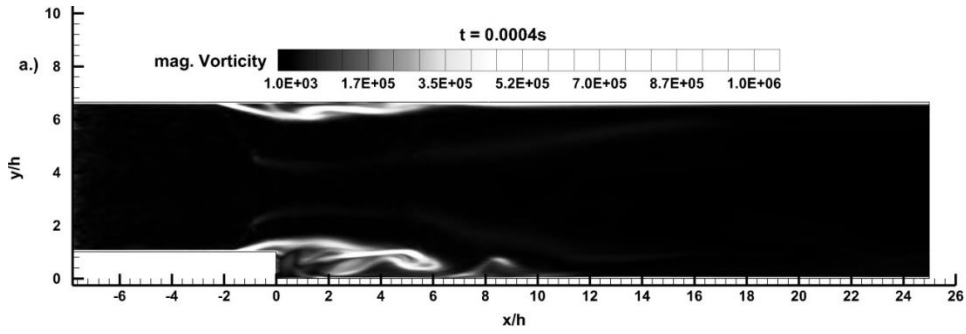

(a) Breakdown of initial vortex with presence of large scale structure

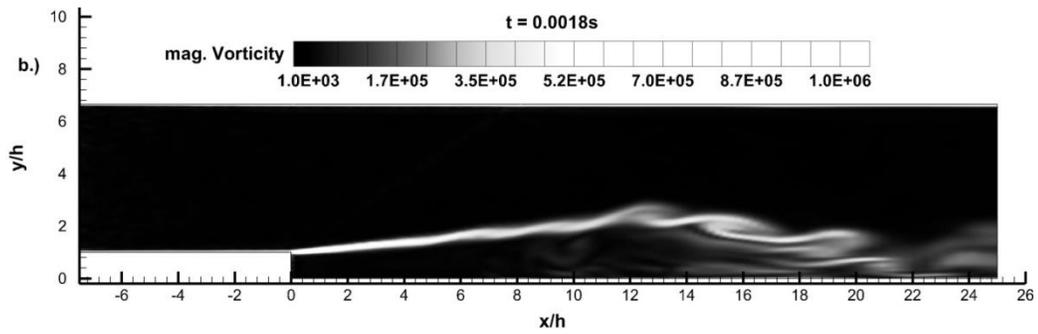

(b) Shear layer starts to roll up

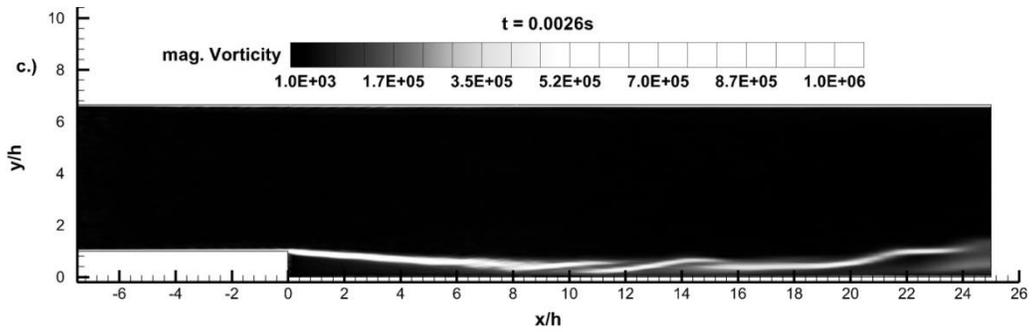

(c) Formation of counter rotating vortex pair

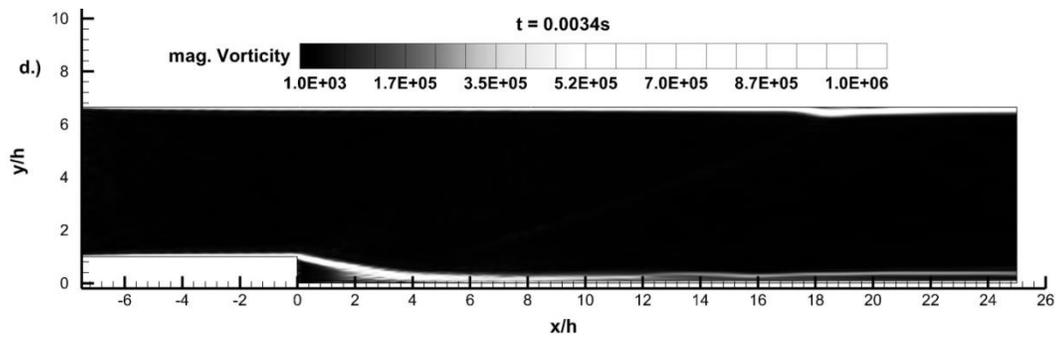

(d) Re-attachment of the shear layer

**Figure 8:** Modulus of vorticity along the centerline ($z=0$) at various instants

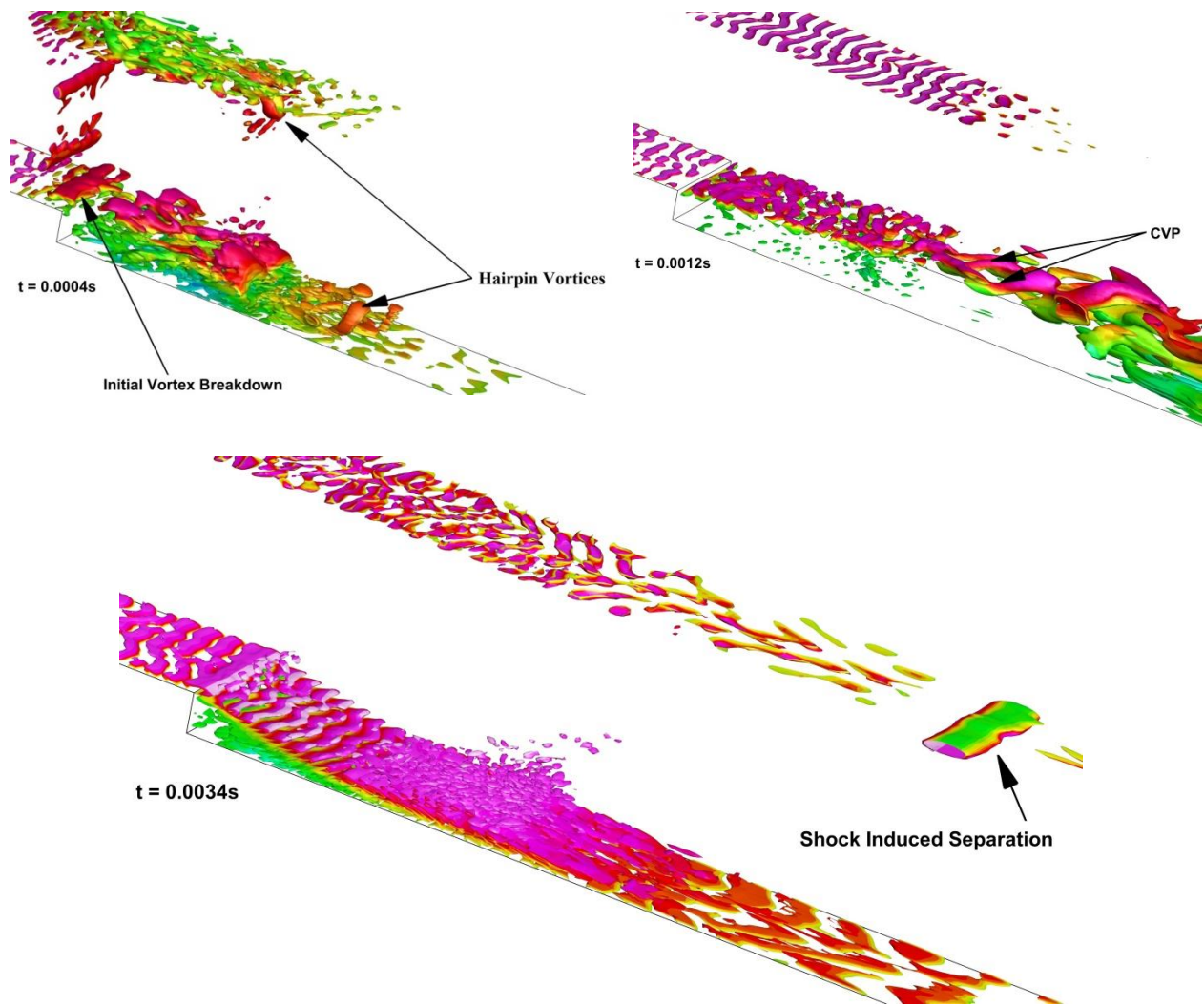

**Figure 9:** Isosurfaces of Q-criterion at various instants depicting various structures in the flow

To identify the coherent structures in the flow field Q-criterion is employed. The dominating structures are visualized by plotting the Isosurfaces of Q-criterion. Isosurfaces at different instants are plotted to understand the evolution of the flow. Q-criterion allows the extraction of 3-Dimensionl vortical structures from the computational data. The Q-criterion deals with the second invariant of the velocity gradient tensor [39]. The Q-criterion is defined as,

$$\left.\begin{aligned} Q &= \frac{1}{2}\left(\Omega_{ij}\Omega_{ij} - S_{ij}S_{ij}\right) \\ \Omega_{ij} &= \frac{1}{2}\left(u_{ij} - u_{ji}\right) \\ S_{ij} &= \frac{1}{2}\left(u_{ij} + u_{ji}\right) \end{aligned}\right\} \quad (22)$$

where $\Omega_{ij}$ and $S_{ij}$ are the anti-symmetric and symmetric part of the velocity gradient tensor $u_{ij}$. To identify the vortex surface, positive Q is plotted. The positive value of Q corresponds to the areas where rotation dominates over plane strain. Also, the vorticity increases towards the center of the vortex core, therefore positive Q values are the good indicator of coherent structures. In Figure 9, Q is plotted to visualize the coherent structure; vortex shedding can be clearly seen which can be attributed to the K-H instability past the step. It is worth noting that the longitudinal vortices are present throughout all the time instants. The hairpin vortices formed close to the step ($x/h \approx$ 1-3), get carried away upon impinging on the lower wall.

After the flow reattaches at around $t \approx$ 0.0034s in the Figure 10, another interesting feature of the flow physics is revealed, widely known as shock wave boundary layer interaction (SWBLI). It can be clearly seen that due to the impingement of the reattachment shock on the upper wall, the boundary layer separates with the formation of smaller re-circulation zone. The noteworthy feature of such interactions, as reported in literature [40, 41], is the vortex shedding beyond the impingement point which is clearly visible at $t \approx$ 0.0034s. The separation from the upper wall has also been reported for incompressible backward facing step flow. Armaly et al. [42] first observed this phenomenon in their experimental study for the Reynolds number range 400-6000 due to adverse pressure gradient. They also observed that beyond Reynolds number 6000 the separation from the upper wall disappeared. However, in the present investigation, Reynolds number being high emough, the separation at the upper wall is primarily due to the shock impingement which is well documented in the shock wave/boundary layer literature. In the Figure 10, $\lambda_2$ and Q-criterion are presented for both the grids after the flow is well established. It is noteworthy that both the vortex identification methodologies exhibit similar flow features on different grids. Also it can be observed that the spanwise roller (at the step corner) stretches and starts to pair again and gets carried away as an arch like structure upon arriving at the wall. This observation is very similar to the helical pairing phenomenon in incompressible [43] flow past backward facing step, however no clear evidence of helical pairing is available in open literature for compressible cases so far.

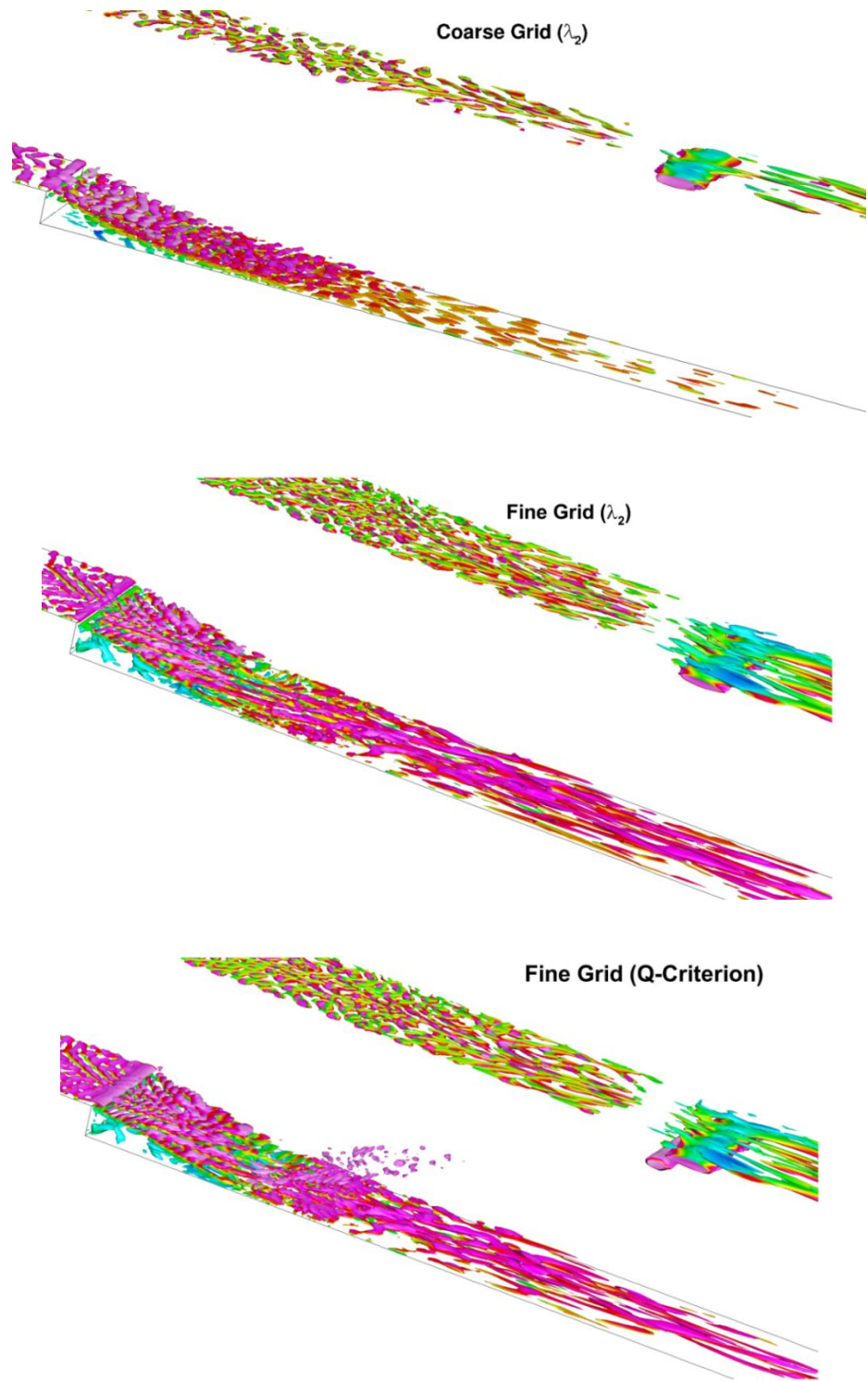

**Figure 10**: $\lambda_2$ and Q-criterion for coarse and fine grid at t ≈ 0.0034s exhibiting similar flow features

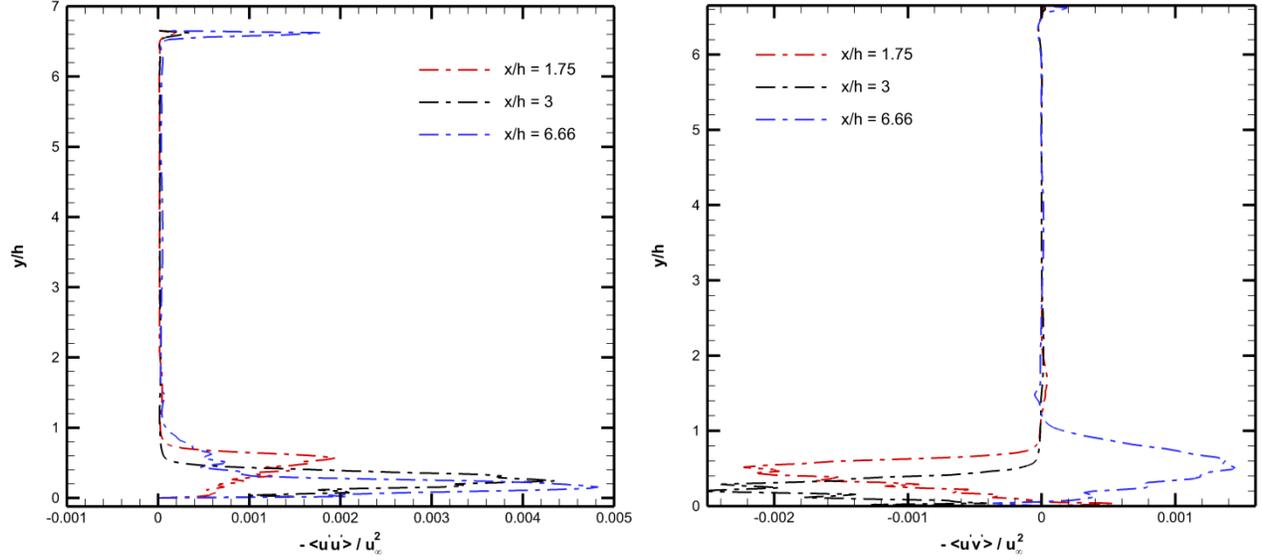

**Figure 11**: Normal and cross stress components at three different locations

The normal and the cross stress components normalized with the streamwise velocity at three different locations are presented in the Figure 11. It can be easily observed that the maximum normal stress component is least in the region near the step and progressively increases downstream of the flow. The cross stress component is positive along $x/h = 6.66$ and negative for the other two locations. Thus, the stress components exhibit the similar flow features as observed in the unsteady data depicted in Figs. [6-9]. Moreover, it is noted that both the coarse and fine grid predictions are in excellent agreement with each other and with measurements as well, while showing exactly similar flow features. Hence, the POD analysis in the next section is carried out using the coarse mesh data only.

**Table 2:** Convergence of first five relative eigenvalue for different number of snapshots

| Mode Number | Relative Eigen Value ($\lambda_i$) | N = 60 | N = 80 | N = 100 | N = 120 |
|---|---|---|---|---|---|
| 1 | $\lambda_1$ | 0.50 | 0.55 | 0.768 | 0.77 |
| 2 | $\lambda_2$ | 0.28 | 0.29 | 0.30 | 0.30 |
| 3 | $\lambda_3$ | 0.069 | 0.062 | 0.059 | 0.059 |
| 4 | $\lambda_4$ | 0.022 | 0.02 | 0.019 | 0.02 |
| 5 | $\lambda_5$ | 0.015 | 0.015 | 0.014 | 0.014 |

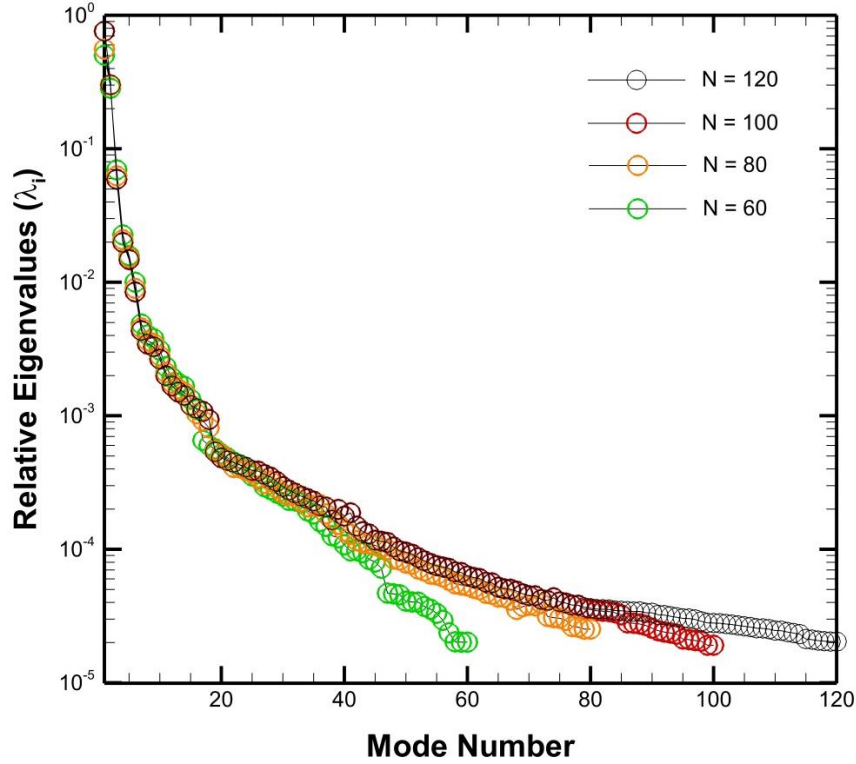

**Figure 12**: Relative Eigenvalue ($\lambda_i$) spectra for different number of snapshots (N)

**3.2 POD Analysis**

Apart from the mean data plot as discussed earlier, POD analysis is performed to capture the coherent structures in the flow field. In order to establish the snap-shots independence, the data for POD analysis is extracted along the centerline plane (Z=0) without any interpolation, i.e. the POD analysis is carried out in the computational grid along different 2D planes (stream-wise & span-wise). Four different snap-shots (snap-shots are collected at a fixed time interval of $10^{-4}$s ) are considered to study the effects of varying snap-shots on the POD results. From Figure 12, it is evident that the Eigen values obtained from the mentioned snap-shots ($\geq$100) seem to have a little difference and they seem to converge within first few modes. Again, Table 2 suggests that the first and second choice for the snap-shots seem to cause a little discrepancy in the Eigen values (1$^{st}$ mode captures only 50-55% of energy) whereas the third and the fourth choice have rendered Eigen values which are in better agreement with each other, and the 1$^{st}$ mode captures almost 77% of energy. Hence, for this current study 100 snap-shots have been selected for detailed analysis.

The POD basis functions for the present case has four components, three corresponding to the velocity and one component corresponding to the temperature. The components corresponding to the streamwise and wall normal velocities, $\phi_x$ and $\phi_y$ (in-plane components) are plotted as vectors and spanwise component $\phi_z$ is presented as contour in the same plot. The length of the vector here has no physical significance unless combined with the coefficients utilized in the reconstruction of the snapshots. The relative eigenvalue can be thought of as the probability of finding a particular coherent structure in a flow field. The mode which has the highest eigenvalue corresponds to the structure which is highly likely to be found in the flow field. Thus, the probability of finding the structure captured in fourth mode is less than the probability of finding the structure in third mode and so on. It is to be noted that only the functions for which the relative eigenvalues are more than 1% of the energy of the most energetic mode contribute to the large scale structures. The in-plane components clearly show a vortical region near the step wall (Figure 13(a)) which is slightly large when compared to re-circulation region as reported in Figure 7(b). It is worth noticing that this vortical region can be seen through all four modes because most of the energy is associated with this dominating structure.

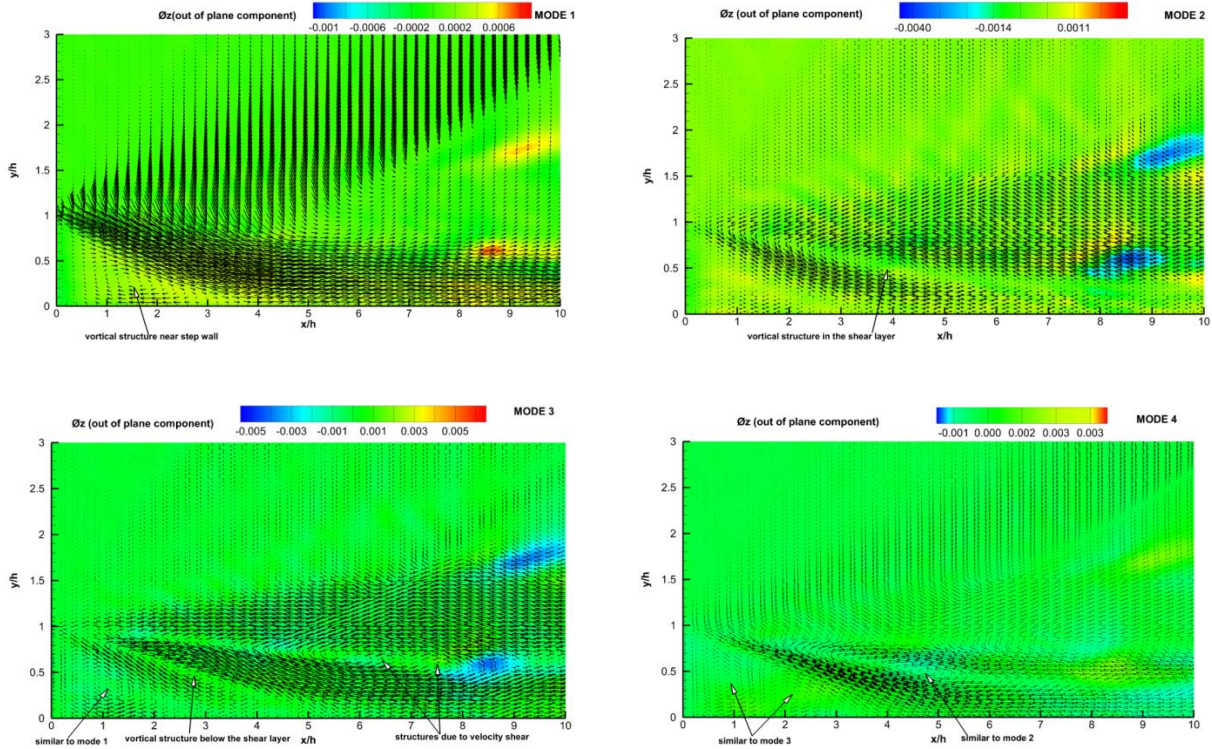

**Figure 13**: Structures associated with different POD modes

The out of plane component has a comparatively high value near the recirculation zone. The second mode (Figure 13(b)) shows vortical structures in the shear layer region with vortex centered at $x/h \approx 4$. The third mode (Figure 13(c)) is quite similar to the second mode; however two vortices appear on either side of the shear layer both centered at $x/h \approx 3$ and $x/h \approx 6$. When comparing this observation with the second mode, it suggests the shifting of the vortex, observed in the second mode at further downstream. The vortical structure near $x/h \approx 6$ is accompanied by another overlapping structure at the downstream. The formation of such structures is attributed to the velocity shear. It can be inferred that these smaller vortices are distributed along the shear layer on the either side of it which is consistent with the observation in modulus of vorticity contour (Figure 8). Similar structure as in mode 2 & 3 can be observed in mode 4 (Figure 13(d)). The energy associated in the modes 3 and 4 is significantly lesser compared to the mode 1. It can be justified by considering the fact that in the modes 3 and 4, the fluctuations are concentrated mainly in the smaller area along the shear layer as opposed to the mode 1 which is associated with the high energy region representing the larger coherent structure. The phase plots of temporal coefficients corresponding to the first and second POD modes and first and third POD modes are presented in Figures 14(a) and 14(b), respectively. The presence of a low dimensional attractor is clearly visible from the plots. Hence, it can be inferred that the decomposition does its job to perform a linear mapping of the data to a lower dimensional space such that the variance of the data in the lower dimensional space is maximized. It is quite apparent that the vortical structures near the step wall and in the shear layer are the most probable coherent structures in the flow.

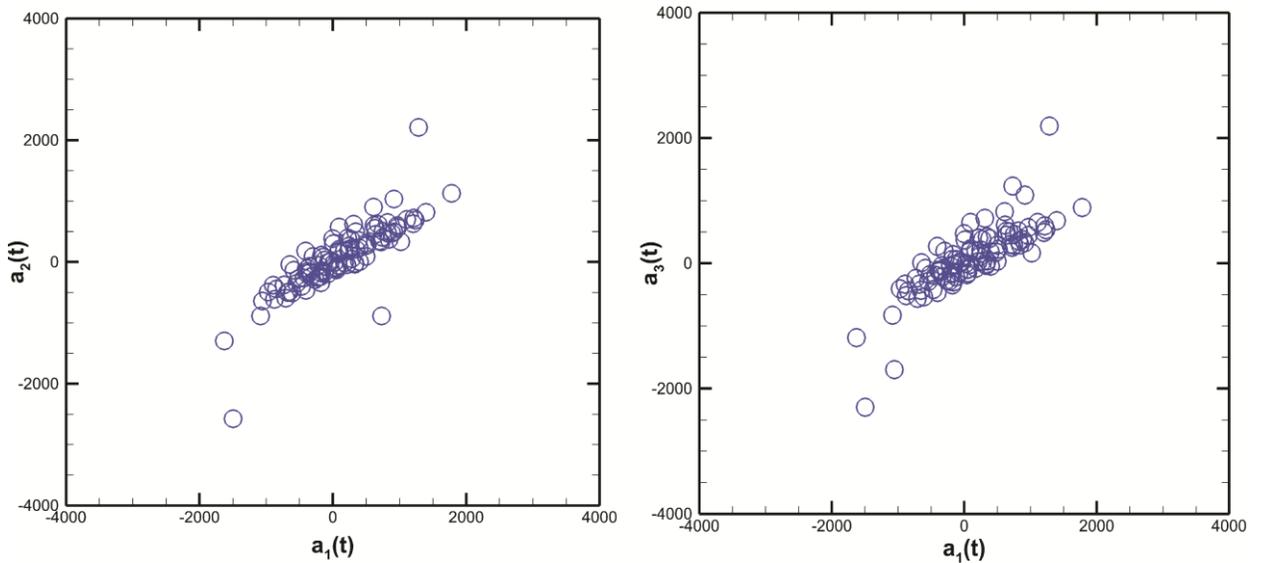

**Figure 14**: Phase plots of temporal coefficients

One of the major applications of POD analysis is the ability to represent the solution with lesser number of POD modes. Figure 15 depicts the reconstruction (Eq. 20) of mean streamwise velocity with 2 & 4 POD modes; with 2 modes few flow features are resolved, however the recirculation region and re-attachment appears to be completely unresolved. This suggests that higher number of modes is required for better representation of the flow field. It can be seen that with only 4 modes all the flow features are resolved accurately, also the solution is reconstructed with 6 & 10 POD modes but the representation is similar to the one with 4 modes and hence it is not presented here.

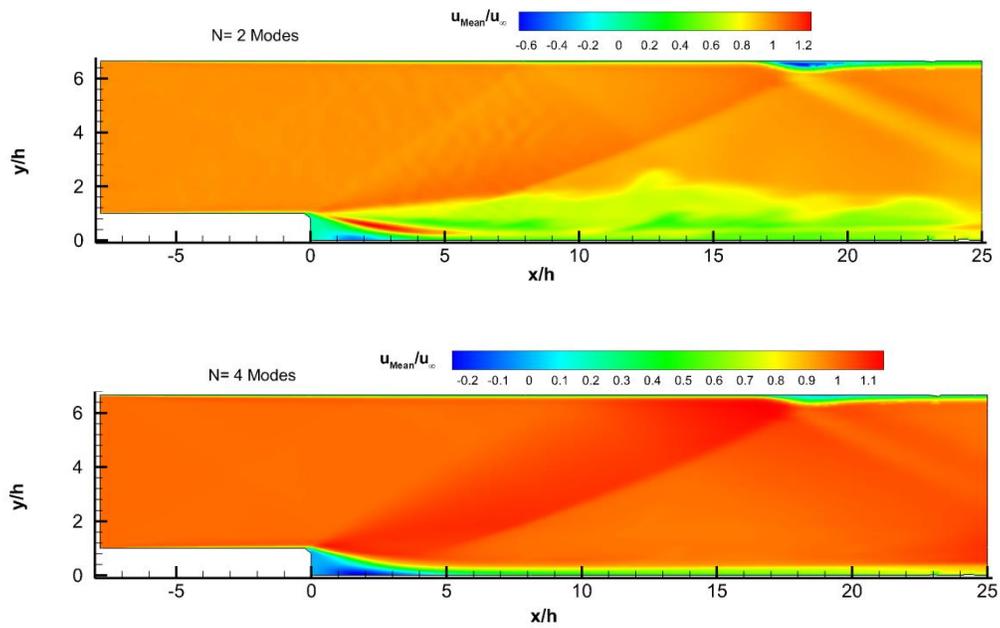

**Figure 15**: Reconstructed contour plot along centerline plane (*xy*) with 2 & 4 POD modes

Apart from the contour plots, quantitative plot of computed mean streamwise velocity at three different locations are compared with the one reconstructed with various POD modes. Figure 16 presents the reconstructed velocity from N = 2, 4, 6 & 10 modes, compared to the computed profile. It is very much interesting to note that with only first four modes the velocity can be represented accurately, even with first 2 modes the velocity away from wall is predicted accurately with discrepancy arising only in the region prone to discontinuity for all *x/h* locations.

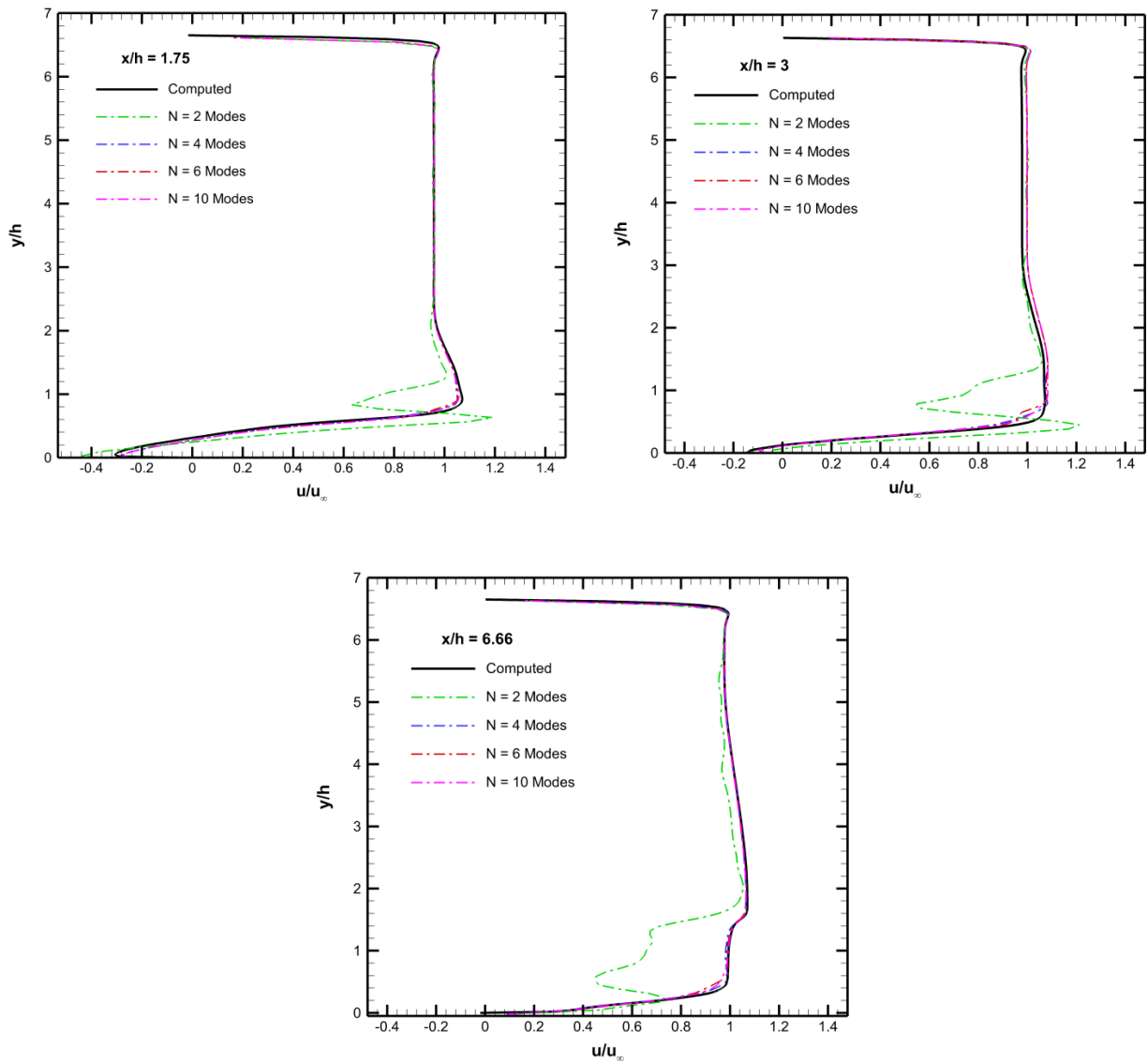

**Figure 16**: Mean streamwise velocity compared with reconstructed velocity

### **POD at *x/h* = 1.75 plane:**

In this section POD results of yz plane at x/h = 1.75 are presented. Figure 17 presents the mean streamwise velocity with in-plane mean (time averaged) and instantaneous velocity vectors. Figure 17(b) suggests the presence of various vortical structures of different sizes, especially in the y/h = 2.5 – 6.6 region with some vortical structures present in near wall region located around *z/h* = 0. Along *y/h* = 3.6, the distribution of in-plane mean vectors appear to be symmetric except slight asymmetry in the near wall region, especially towards the lower wall. This is due to the fact that the lower wall is located within the recirculation region giving rise to this asymmetry.

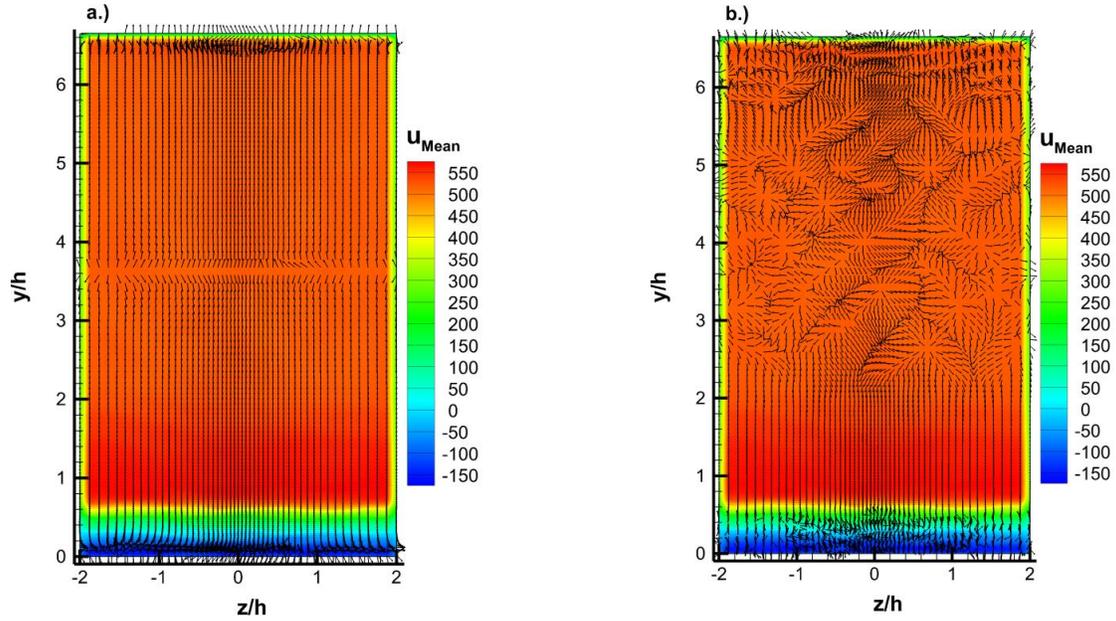

**Figure 17**: a) Mean out of-plane velocity with mean in-plane vectors and b) Mean out of-plane velocity contour with instantaneous in-plane vector

In the Figure 18, first 4 POD modes are shown for *x/h* = 1.75 plane, and it is observed that in 1st mode high value of $\varphi_x$ is distributed throughout the plane with presence of some vortical structures around *y/h* = 3 & 4. The first mode alone represents 96 % of energy of velocity fluctuations and closely mimics the mean flow behavior; which can be verified on comparing it with Figure 17(a). Mode 2 represents only 0.3 % energy and shows large cluster of vortical structures between y/h =2 & 6.6, including the presence of some vortical structures close to the wall. Similar kind of behavior especially in the upper part of the plane is noticed for 3rd and 4th modes as well. However it is interesting to note that with decreasing energy level more vortical structures become apparent, especially near the bottom wall region. Pedersen and Meyer [44] also came across similar observation in their study and they found that for 200th mode more details were present.

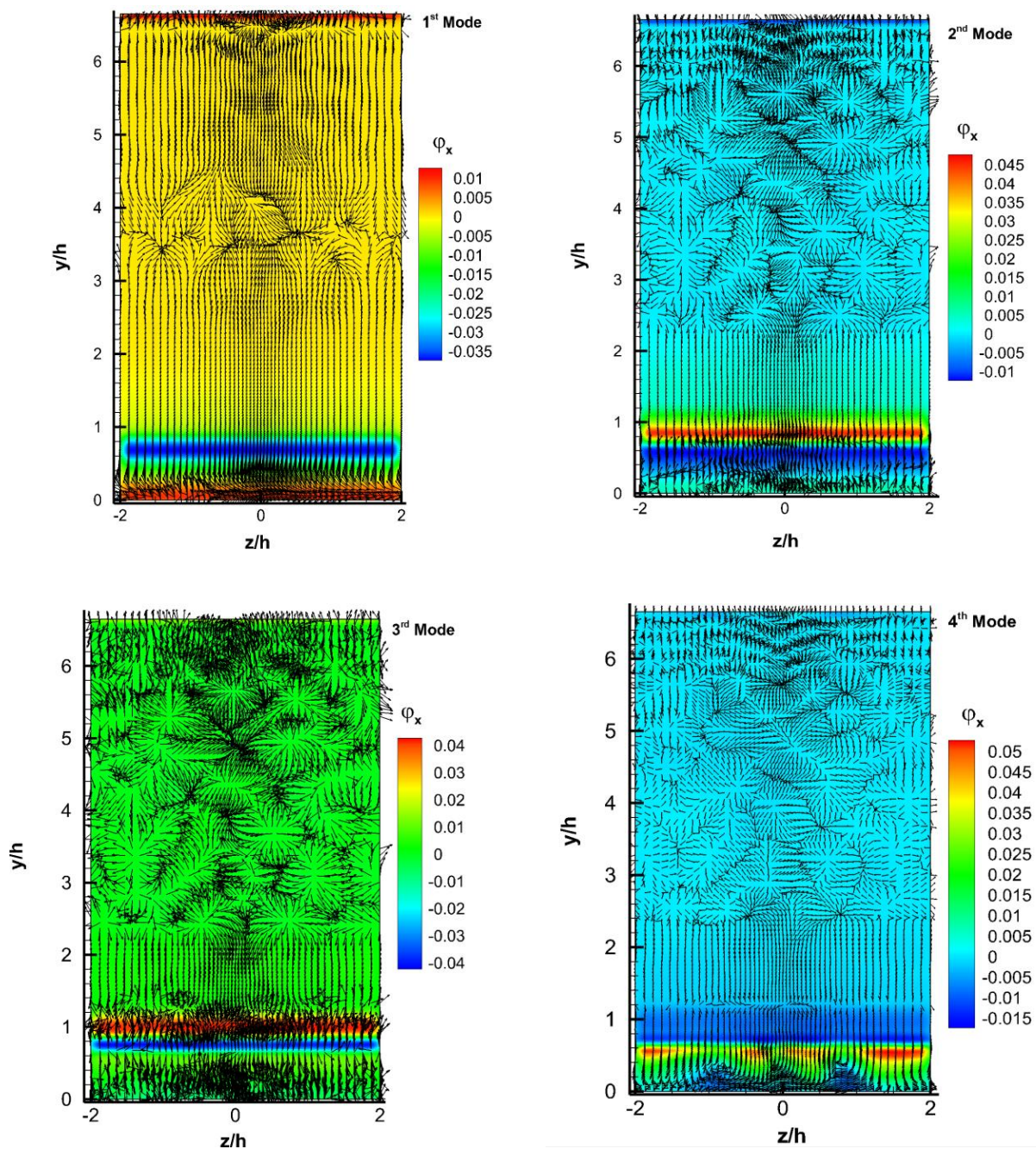

**Figure 18**: 1st four POD modes for x/h = 1.75 plane (Contour: Out of –plane component)

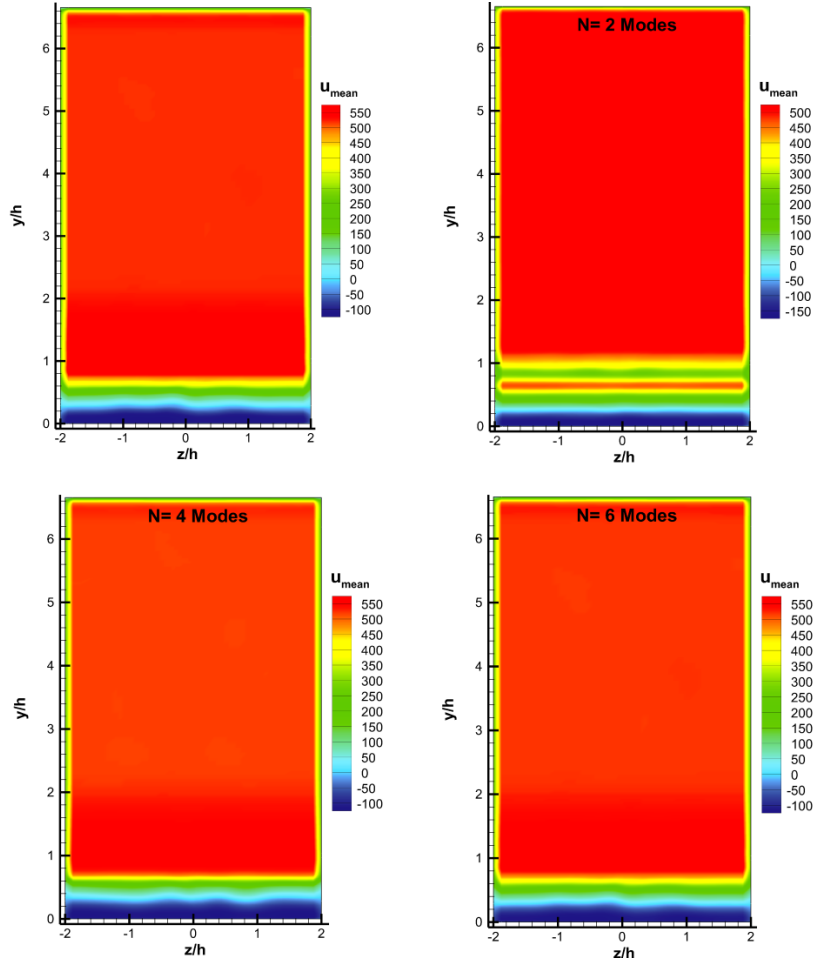

**Figure 19**: Mean streamwise velocity and reconstructed contour plot along x/h = 1.75 with N= 2, 4 & 6 POD modes

As the presence of vortices close to the bottom wall is evident; these vortical structures appear similar to the counter rotating vortex pair which occurs during the evolution of the flow field. The energy represented by $3^{rd}$ and $4^{th}$ mode is approximately 0.038 % & 0.0077 % respectively. It is worth noticing that the representation from $2^{nd}$, $3^{rd}$ & $4^{th}$ mode follow the instantaneous behavior of flow field which can be verified on comparing it with Figure 17(b). The observation of higher modes cannot be thought of as the presence of dominant flow structures, rather these are the representation of small scale motions.

Figure 19 presents the reconstructed contour of mean streamwise velocity with N = 2, 4 & 6 modes. The reconstruction is performed as described in eqn. 20, again the original flow field is retained with only first four modes whereas with first 2 modes strip of high velocity region about $y/h \approx 0.6$ is present. This observation is consistent with the line plots of mean velocity in Figure 16. Hence it suggests that with only first 4 modes, the complete description of flow field can be extracted. Similarly variables are reconstructed for x/h = 3 & 6.66 plane and it has been noticed that the first 4 modes are sufficient to represent the flow field accurately.

**POD at *x/h = 6.66* plane**

In this section, we present close up view of near wall region for x/h = 6.66 plane in order to complement our discussion related to Figure 18. The $1^{st}$ mode captures approximately 91 % of the energy; higher value of out of plane component is present close to the wall with lesser details relating to the evolution of the flow field (Figure 20). The representation from mode 1 follows the mean flow field closely. Mode 2, which represents 7 % of the total energy, exhibits some small vortical structures but appear to be insignificant. However, mode 3 onwards the details start to increase with the decreasing energy. The presence of counter rotating vortex pair is evident for the $4^{th}$ mode which represents only 0.006 % of the total energy. This suggests that the CVP is a transient structure and is not present once the steady state is reached. Mode 5 and 6 also suggest the presence of some structures, however they are not significant as these structures only represent small scale motions. Mode 5 and 6 capture only 0.004 % and 0.002 % of the total energy.

The PSD of temporal coefficients corresponding to the POD modes obtained in the plane *z=0* is presented in Figure 21. The frequency corresponding to the highest peak for the temporal coefficient corresponding to the mode 1 is 2.125 * 10^5, whereas the peak frequency corresponding to the other three modes is 3.9*10^5 which is close to the peak frequency obtained for the spectra presented in the Figure 4. The first POD mode captures the maximum energy for the present flow and it is clear from the POD modes presented in the Figures 1, 18 & 20 that the first mode is a representation of the mean properties. The first temporal mode closely represents the mean pressure while the pressure fluctuations are represented by the modes 2-4. The location *x/h = 1.75* corresponds to the recirculation region near the step. It is quite clear from the mean pressure contours that the pressure in this region does not exhibit large fluctuations. Therefore, all the modes follow nearly the same characteristics. The location *x/h = 3* also lies in the recirculation region closer to the reattachment point. The mean pressure peaks corresponding to the mode *a1* are higher than that obtained for the plane *x/h=1.75* which is consistent with the observation of the mean pressure contours. It can also be noted that the peaks corresponding to the modes *a2-a4* are also higher than that obtained for the plane at *x/h = 1.75* indicating increased pressure fluctuations. The plane at *x/h = 6.66*, representing region after the reattachment point, displays further increase in the pressure fluctuations with higher peaks shifted towards the higher frequencies.

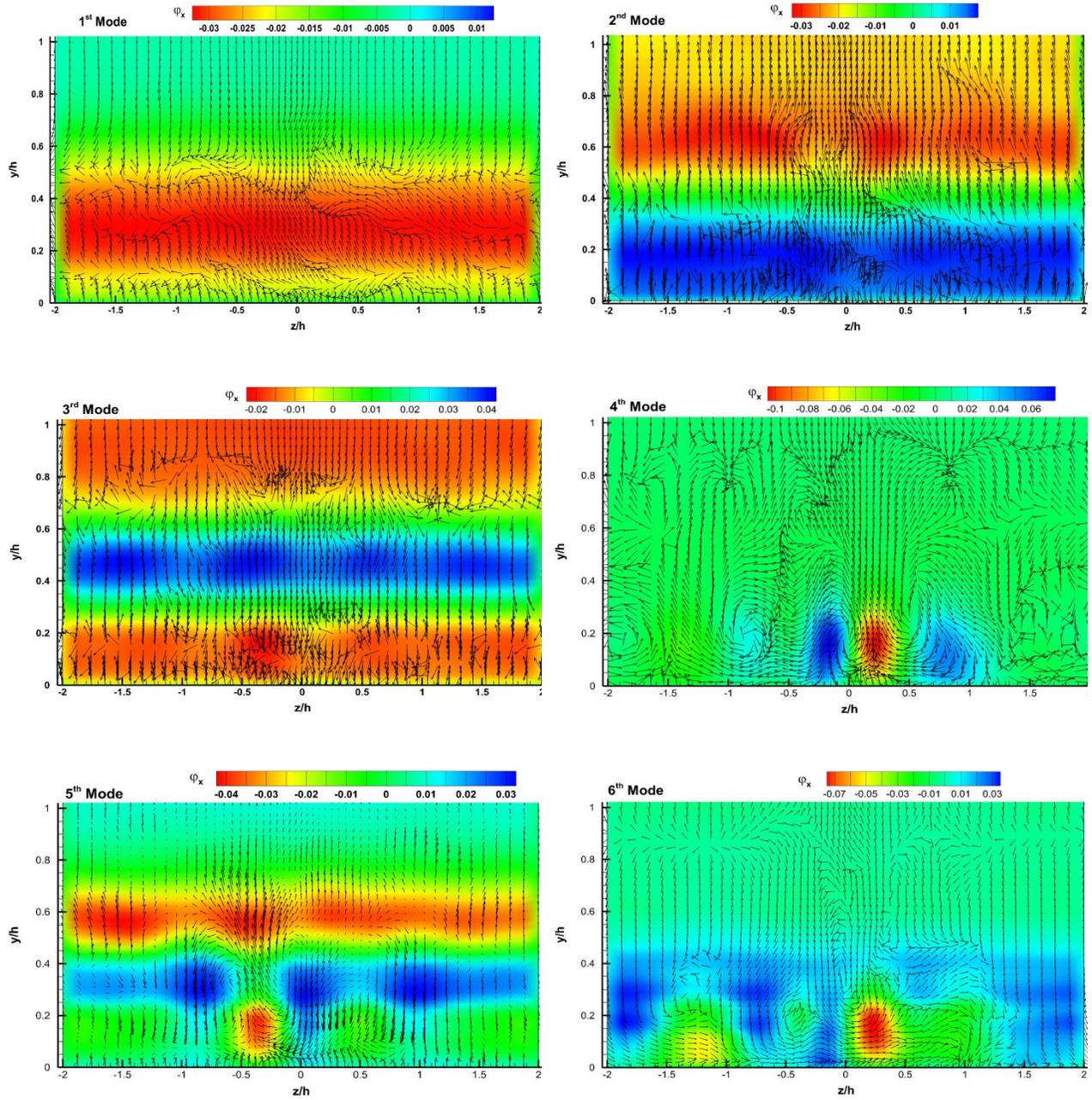

**Figure 20**: Close up view of 1st six POD modes for x/h = 6.66 plane (Contour: Out of plane component)

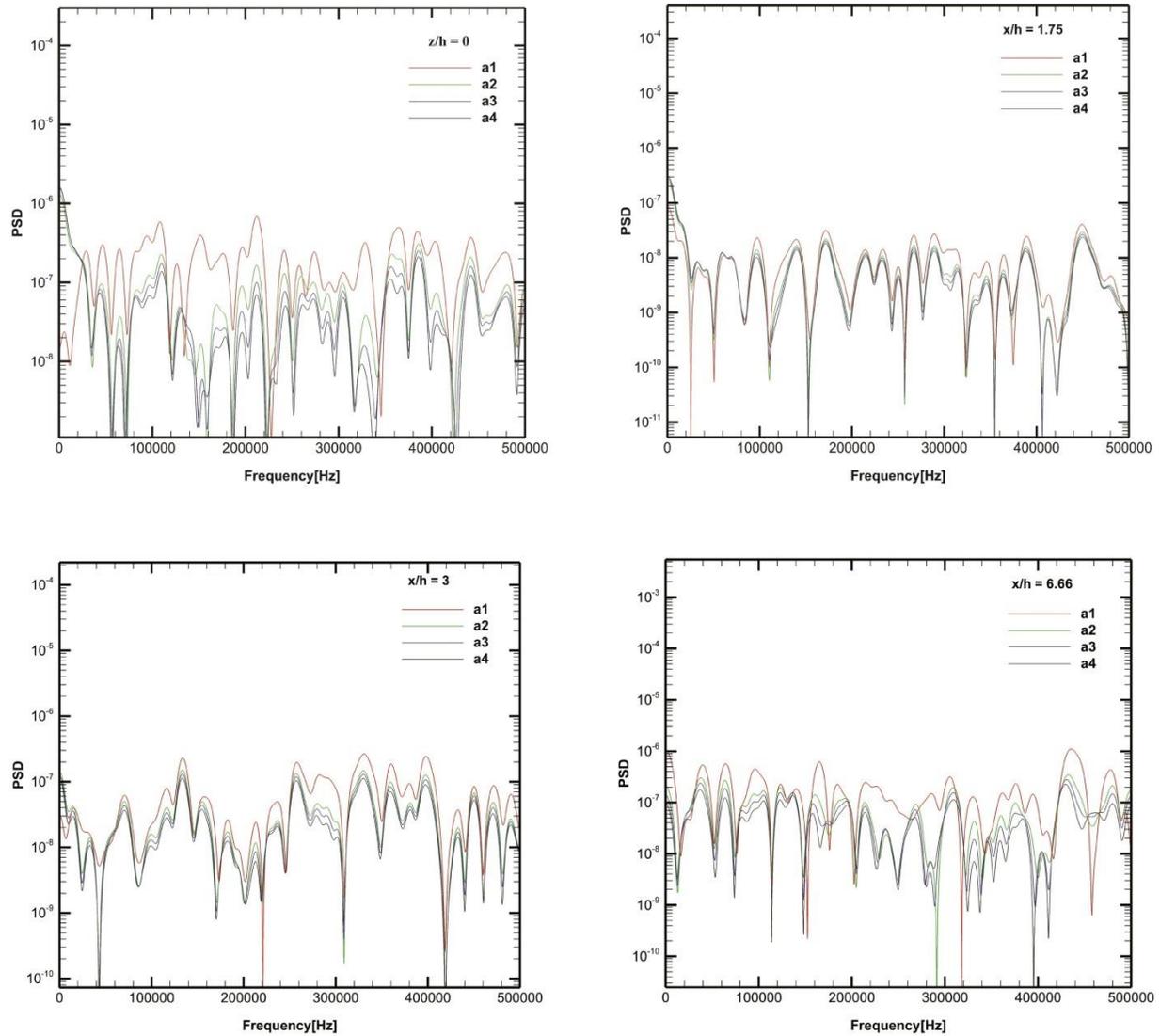

**Figure 21**: PSD of the temporal coefficients along various planes

On combining the observation of POD along with Q and the modulus of vorticity, it is clearly visible that the predicted results complement each other with excellent agreement. As noticed from the Figure 8 that the shear layer tends to oscillate after re-attachment, while the similar phenomenon can be deduced from Figure 9. The vortical shedding is observed clearly for $z = 0$ plane for $2^{nd}$, $3^{rd}$ & $4^{th}$ mode. The study, extended to spanwise planes at two different axial locations, also reveals the various transient flow features similar to those noticed in Figures 8 & 9. The PSD of temporal coefficient for different plane seems to follow the pressure oscillation closely to what is predicted by the Fourier transform of fluctuating pressure.

## 4. CONCLUSIONS

The Large Eddy Simulation of backward facing step at Reynolds number of $1.024 \times 10^5$ and Mach 2 is performed in conjunction with POD analysis. The used wall function modelling is validated against the DNS results of supersonic flow over a flat plate. The results show that a wall model with LES provides an efficient way to reduce the computational cost of a conventional LES without compromising much on the accuracy of the solution. As far as the mean flow features are concerned, the simulations appear to resolve the flow physics with good agreement, except for the region $y/h < 1$. The insight in to the flow physics is gained by utilizing Q-criterion, modulus of vorticity and POD modes. Q and modulus of vorticity plots reveal the SWBLI interaction, formation of counter rotating vortex pairs and the shear layer flapping. Other important observations worth mentioning are shock induced separation leading to vortex shedding and the shear layer instability. The proper orthogonal decomposition performed on the computed data is in agreement with the flow physics put forward through many experimental studies. Also it is in good agreement with the observation from the iso-surfaces of Q and the modulus of vorticity contour plots. First four modes are presented which confirm the presence of re-circulation region and small vortices along the shear layer as the major coherent structures ($z = 0$). In the spanwise plane, the decomposition reveals many structures corresponding to low energy modes. Though, these structures are transient and they vanish once the flow reach a steady state, yet at the same time they provide deep insight into the evolution of the flow field. The representation of these transient structures through low energy modes also complements the fact that coherent structures correspond to the higher energy modes. The PSD of the temporal coefficients corresponding to the first four POD modes along three different planes shows that the POD coefficients are useful in determining the pressure fluctuations, results of which are corroborated by the LES results. Thus, it is shown that the POD modes and the POD coefficients represent essential features of the flow field. The present work successfully combines and correlates the major results provided by LES and subsequent lower order decomposition using POD.

## 5. ACKNOWLEDGEMENTS

Financial support for this research is provided through IITK-Space Technology Cell (STC). Also, the authors would like to acknowledge the IITK computer center (www.iitk.ac.in/cc) for providing the resources to perform the present work.